\documentclass[fleqn]{article}
\usepackage{gc,amsfonts}

\heads{K.A. Bronnikov, G.N. Shikin and E.N. Sibileva }
    {Self-gravitating string-like configurations
        from nonlinear electrodynamics }

\def\sph{spherically symmetric}
\def\ssph{static, spherically symmetric}
\def\cyl{cylindrically symmetric}
\def\asflat{asymptotically flat}

\def\duF{{}^*F}

\def\mn{_{\mu\nu}}
\def\MN{^{\mu\nu}}
\def\mN{_\mu^\nu}
\def\nM{_\nu^\mu}

\def\R{{\mathbb R}}

\def\fpi{\frac{1}{16\pi}}

\def\kappa{\mbox{\sl \ae}}

\begin{document}
\twocolumn[
\prepno{gr-qc/0308002}{}

\Title {Self-gravitating string-like configurations\yy
    from nonlinear electrodynamics }

\Authors{K.A. Bronnikov} {G.N. Shikin and E.N. Sibileva}
    {VNIIMS, 3-1 M. Ulyanovoy St., Moscow 117313, Russia;\\
        Institute of Gravitation and Cosmology, PFUR,
        6 Miklukho-Maklaya St., Moscow 117198, Russia}
    {Department of Theoretical Physics, PFUR,
        6 Miklukho-Maklaya St., Moscow 117198, Russia}

\Abstract
{We consider static, cylindrically symmetric configurations in general
relativity coupled to nonlinear electrodynamics (NED) with an arbitrary
gauge-invariant Lagrangian of the form $L_{\rm em}= \Phi(F)$, $F =
F_{\alpha \beta} F^{\alpha \beta}$. We study electric and magnetic fields
with three possible orientations: radial (R), longitudinal (L) and azimuthal
(A), and try to find solitonic stringlike solutions, having a regular axis
and a flat metric at large $r$, with a possible angular defect. Assuming that
the function $\Phi(F)$ is regular at small $F$, it is shown that a regular
axis is impossible in R-fields if there is a nonzero effective electric
charge and in A-fields if there is a nonzero effective electric current along
the axis. Thus solitonic solutions are only possible for purely magnetic
R-fields and purely electric A-fields, in cases when $\Phi(F)$ tends to a
finite limit at large $F$. For both R- and A-fields it is shown that the
desired large $r$ asymptotic is only possible with a non-Maxwell behaviour
of $\Phi(F)$ at small $F$. For L-fields, solutions with a regular axis are
easily obtained (and can be found by quadratures) whereas a desired large $r$
asymptotic is only possible in an exceptional solution; the latter gives rise
to solitonic configurations in case $\Phi(F) = \const \cdot \sqrt{F}$. We
give an explicit example of such a solution.}

] %%%%%%%%%%%%%%%%%%%%%%%%%%%%%%%%%%%%%%

\section{Introduction}

    Self-gravitating systems with classical nonlinear field sources find
    many applications in modern unification theory and cosmology.
    On the one hand, there naturally appear various scalar fields with
    nonlinear potentials and different coupling to gravity, on the other,
    some models of string theory create in their low-energy limits such
    theories as the Born-Infeld nonlinear electrodynamics or its non-Abelian
    modifications \cite{BI}. Of utmost importance in any such theory are
    solitonic or solitonlike solutions, combining global regularity and
    localization in a certain number of dimensions, such as particlelike
    solutions, kinks, domain walls, cosmic strings, etc.

    Of particular interest are cylindrically symmetric (stringlike)
    solutions, with fields localized in a vicinity of the symmetry axis.
    Nielsen and Olesen \cite{Nielsen} pointed out that vortex solutions of
    the Higgs model behave classically as Nambu dual strings \cite{Nambu} in
    the strong-coupling limit. As an example, they have shown that a
    Higgs-type Lagrangian allows for vortex-lines solutions, similar to
    vortex lines in a type II superconductor. Self-gravitating cosmic
    strings have been a subject of great interest in cosmology in the
    recent decades, see Refs.\,\cite{vilenk,meier} for reviews.

    Stringlike configurations of nonlinear fields have also been studied: in
    \Ref {Bron2} for self-interacting scalar fields and in \Ref{Shikin84}
    for interacting scalar and electromagnetic fields.

    The present paper continues this trend and studies the possibility of
    obtaining regular self-gravitating stringlike configurations of a
    nonlinear electromagnetic field with an arbitrary gauge-invariant
    Lagrangian $\Phi(F)$, $F=F\mn F\MN$. We consider three possible types of
    electromagnetic fields compatible with cylindrical symmetry: radial (R),
    azimuthal (A) and longitudinal (L) fields. We formulate the conditions
    of the solutions' regularity on the symmetry axis and for their proper
    behaviour far from the axis and conclude that solitonic stringlike
    configurations, i.e., those well-behaved both on the axis and at
    infinity, can be built in the cases of R- and A-fields under certain
    conditions upon the function $\Phi(F)$. The regular axis conditions are
    somewhat similar to the regular centre conditions for the NED-Einstein
    system in the case of spherical symmetry \cite{Bron1}.

    According to \cite{Bron1}, it is possible to obtain regular \asflat\
    purely magnetic solutions (black holes and monopoles) in NED which
    reduces to the Maxwell theory ($\Phi\sim F$) at small $F$. Unlike that,
    regular (flat or string) asymptotics of \cyl\ configurations are only
    possible when $\Phi(F) = o(F)$ at small $F$. This circumstance is
    closely related to the fact that the corresponding Einstein-Maxwell
    solutions are not well-behaved at infinity.

    For L-fields, although a regular axis is easily obtained (even in the
    Einstein-Maxwell system \cite{Melvin,Bron3}), the general solution with
    a regular axis (which is obtained by quadratures) cannot lead to a
    stringlike solitonic configuration since a regular spatial asymptotic is
    absent. There is, however, an exceptional purely magnetic solution with
    $\Phi(F) = \const\cdot \sqrt{F}$ which turns out to be solitonic under
    some further restrictions.

    It should be stressed that, in nonsingular configurations of any spatial
    symmetry, effective electric and magnetic charges, characterizing the
    field behaviour at large (e.g., at a spatial asymptotic), appear
    in nonlinear theory without postulating the existence of such charges in
    the initial formulation of the theory (as is done in the Maxwell theory
    where the current density $j^{\mu}$ is introduced in the interaction
    term $j^{\mu}A_\mu$, added to the field Lagrangian $\sim F\MN F\mn$).
    It is the field nonlinearity that leads to effective electric and
    magnetic charge densities or currents distributed in space. In this way
    magnetic monopoles and regular black holes with magnetic charges were
    obtained in \Ref{Bron1} without introducing a magnetic charge density
    as a separate quantity; their \cyl\ analogues are solitonic
    configurations whose existence is considered here.

\section {Field equations and regularity conditions}  %% s2

\subsection{Field equations}

    Consider gauge-invariant NED in general relativity (GR), with the action
\beq
    S =  \frac{1}{2\kappa}\int \sqrt{-g}\,d^4 x[R- G \Phi(F)],   \label{L}
\eeq
    where $R$ is the scalar curvature, $g$ is the determinant of the metric
    tensor $g\mn$, $G$ is the gravitational constant, $\kappa = 8\pi G$, and
    $\Phi(F)$ is an arbitrary function of the invariant $F = F_{\alpha
    \beta} F^{\alpha \beta}$.  Maxwell's electrodynamics corresponds to
    $\Phi(F) \equiv F$, and, in nonlinear theory, it is natural to assume a
    Maxwell behaviour ($\Phi(F) \approx F$) at small $F$.

    The electromagnetic field equations following from (\ref{L}) and the
    Bianchi identities for $F\mn$ have the form
\bear
    \nabla_\nu (F\MN \Phi_F)\eql 0, \cm  \Phi_F \equiv \frac{d \Phi}{d F},
                                                          \label{eqF}
\\
    \nabla_\nu \duF\MN \eql  0, \quad
          \duF\MN = -\frac{1}{2\sqrt{-g}}
            \eps^{\lambda\rho\mu\nu} F_{\lambda \rho},
                                                   \label{eqF*}
\ear
    where $\eps^{\lambda \rho \mu \nu}$ is the Levi-Civita symbol and $*$
    denotes the Hodge dual.

    Variation of (\ref{L}) with respect to $g\MN$ leads to the Einstein
    equations
\beq
    G\mN \equiv R\mN - \half \delta\mN R = - \kappa T\mN        \label{EE}
\eeq
    with the electromagnetic stress-energy tensor (SET)
\beq
    T\mN = \frac{1}{16\pi}[-4F_{\mu\alpha}F^{\nu\alpha}\Phi_F
                            + \delta\mN \Phi]. \label{SET}
\eeq

    A static, cylindrically symmetric metric can be written as
\beq
    ds^2 = \e^{2\gamma} d t^2 - \e^{2\alpha} d u^2 -
            \e^{2\xi} d z^2 - \e^{2\beta} d \varphi ^2,       \label{ds}
\eeq
    where $\alpha,\ \beta,\ \gamma,\ \xi$  are functions of the radial
    coordinate $u$ only; $z \in \R$ and $\varphi \in [0;2\pi)$ are the
    longitudinal and azimuthal coordinates, respectively. There is still
    freedom of choosing the $u$ coordinate.

    We find it convenient to use the coordinate condition \cite{Bron3}
\beq
    \alpha = \gamma + \xi + \beta,                           \label{harm}
\eeq
    so that $u$ is a harmonic coordinate. Its range is not specified until
    the full geometry is known. Under the condition (\ref{harm}), \eqs
    (\ref{EE}) take the following symmetric form:
\bear
    \beta'' + \xi''- U \eql - \kappa\, T_0^0 \e^{2\alpha},  \label{00}
\\
    U \eql - \kappa {T_1 ^1} \e^{2\alpha},              \label{11}
\\
    \beta'' + \gamma''- U \eql - \kappa\, T_2^2 \e^{2\alpha},\label{22}
\\
    \gamma'' + \xi''- U
             \eql - \kappa\, {T_3 ^3} \e^{2\alpha},         \label{33}
\\
    U \al\equiv\al \beta'\gamma' + \beta'\xi' + \gamma'\xi'.
\earn

    We will seek solutions in three cases compatible with cylindrical
    symmetry, with the following nonzero components of $F\mn$ in each case:
\begin{description}  \itemsep 0pt
\item[] \nhq
      Radial (R) fields: electric, $F_{01} (u)$ ($E^2 =F_{01}F^{10}$), and
      magnetic, $F_{23} (u)$ ($B^2 = F_{23}F^{23}$).
\item[] \nhq
      Azimuthal (A) fields: electric, $F_{03} (u)$ ($E^2 =F_{03}F^{30}$),
      and magnetic, $F_{12}(u)$ ($B^2 = F_{12} F^{12}$).
\item[] \nhq
      Longitudinal (L) fields: electric, $F_{02}(u)$  ($E^2
      =F_{02}F^{20}$), and magnetic, $F_{13}(u)$ ($B^2 = F_{13}F^{13}$).
\end{description}

    Here $E$ and $B$ are the absolute values of the electric field strength
    and magnetic induction, respectively.

\subsection{Regularity on the axis and regular (flat or string) spatial
        asymptotics}

    From the whole set of solutions to the field equations, we will try to
    single out those having (i) a regular axis of symmetry and (ii) such a
    behaviour far from the axis that corresponds to the gravitational field
    of an isolated \cyl\ matter distribution or a cosmic string, i.e., a
    flat or string asymptotic at spatial infinity (in what follows, a {\it
    regular\/} asymptotic, for short).

    Let us first write the regularity conditions without specifying the $u$
    coordinate.

    The regularity conditions on the axis, i.e. for $u = u_a $ such that
    $r \equiv \e^{\beta}\to 0$, include the finiteness requirement
    for the algebraic curvature invariants and the condition
\beq
    |\beta'| \e^{\beta - \alpha} \to 1                  \label{circ}
\eeq
    (the prime denotes $d/du$), expressing the absence of a conical
    singularity.

    Among the curvature invariants it is sufficient to deal with the
    Kretschmann scalar $K = R^{\mu\nu\rho\sigma}R_{\mu\nu\rho\sigma}$ which,
    for the metric (\ref{ds}), is a sum of all squared nonzero Riemann tensor
    components $R\MN{}_{\rho\sigma}$:
\bear
    K \eql 4 \sum_{i=1}^{6} K_i^2;
\nnv
    K_1 \eql R^{01}{}_{01}
            =-\e^{-\alpha-\gamma}(\gamma'\e^{\gamma-\alpha})',
\nn
    K_2 \eql R^{02}{}_{02} = -\e^{-2\alpha}\gamma'\xi',
\nn
    K_3 \eql R^{03}{}_{03} = -\e^{-2\alpha}\beta'\gamma',
\nn
    K_4 \eql R^{12}{}_{12} = -\e^{-\alpha-\xi}(\xi'\e^{\xi-\alpha})',
\nn
    K_5 \eql R^{13}{}_{13} = -\e^{-\alpha-\beta}(\beta'\e^{\beta-\alpha})',
\nn
    K_6 \eql R^{23}{}_{23} = -\e^{-2\alpha}\beta'\xi'            \label{Kr}
\ear
    For $K < \infty$ it is thus necessary and sufficient that all $|K_i| <
    \infty$, and this in turn guarantees that all algebraic invariants of
    the Riemann tensor will be finite. Note that all $K_i$, as well as the
    condition (\ref{circ}), are invariant under reparametrization of $u$.

    It can be verified (see \cite{Bron2} for details) that the regular axis
    conditions hold at $u=u_{\rm ax}$, where $r = \e^{\beta} \to 0$, if and
    only if
\bearr
     \gamma = \gamma_{\rm ax} + O(r^2); \cm
                 \xi = \xi_{\rm ax} + O(r^2),       \label{ax1}
\yyy
     |\beta'| \e^{\beta-\alpha}= 1 + O(r^2)                 \label{ax2}
\ear
    as $u \to u_{\rm ax}$. Here and henceforth the symbol $O(f)$ denotes a
    quantity either of the same order of magnitude as $f$ in a certain limit,
    or smaller, while the symbol $\sim$ connects quantities of the same order
    of magnitude.

    A useful necessary condition for regularity follows from the Einstein
    equations. At points of a regular axis, as at any regular point, the
    curvature invariants $R$ and $R\mn R\MN$ are finite. Since the Ricci
    tensor for the metric (\ref{ds}) is diagonal, the invariant $R\mn R\MN
    \equiv R\mN R\nM$ is a sum of squares, hence each component $R_\mu^\mu$
    (no summing) is finite. Then, due to the Einstein equations, each
    component of the SET $T\mN$ is finite as well:
\beq
    |T\mN| < \infty.                                     \label{Tfin}
\eeq
    Thus, requiring only regularity of the geometry, we obtain, as its
    necessary condition, the finiteness of all SET components with mixed
    indices. This is true not only for the present case, but always when
    $R\mN$ is diagonal.

    Let us now formulate the conditions at regular spatial asymptotics.
    We require the existence of a spatial infinity, i.e., $u = u_\infty$
    such that $r = \e^\beta \to \infty$, where the metric is either flat, or
    corresponds to the gravitational field of a cosmic string.

    Then, first, as $u\to u_\infty$, a correct behaviour of clocks and
    rulers requires $|\gamma| < \infty$ and $|\xi|<\infty$ or, choosing
    proper scales along the $t$ and $z$ axes, one can write
\beq
    \gamma\to 0,\ \ \ \xi\to 0 \ \ \
    {\rm as} \ \ u \to u_\infty.                             \label{as1}
\eeq

    Second, at large $r$ the condition (\ref{circ}) should be replaced
    with a more general one,
\bearr
    |\beta'|\e^{\beta-\alpha} \to 1 - \mu,
\nnn
    \mu = \const < 1, \ \ \ {\rm as} \ \ \ u\to u_\infty,       \label{mu}
\ear
    so that the circumference-to-radius ratio for the circles $u=\const$,
    $z=\const$ tends to $2\pi (1-\mu)$ instead of $2\pi$, $\mu$ being the
    angular defect. Under the asymptotic conditions (\ref{as1}),
    (\ref{mu}), $\mu>0$, the solution can simulate a cosmic string. A flat
    asymptotic takes place if $\mu = 0$.

    Third, the curvature tensor should vanish at large $r$, and, due to
    the Einstein equations, all SET components must decay quickly enough.
    It can be easily checked that the conditions (\ref{as1}) and (\ref{mu})
    automatically imply that all $K_i = o(r^{-2})$ where $K_i$ are defined
    in (\ref{Kr}). Consequently the same decay rate at a regular asymptotic
    takes place in all components of $T\mN$, and one can verify, in
    particular, that the total material field energy per unit length along
    the $z$ axis is finite:
\beq
    \int T^0_0 \sqrt{-{}^3g}\, d^3x =
    \int T^0_0 \e^{\alpha+\beta+\xi}\,du\,dz\,d\phi < \infty  \label{Efin}
\eeq
    where integration in $z$ covers a unit interval. A similar condition in
    flat-space field theory is used as a criterion of field energy being
    localized around the symmetry axis, which is one of the requirements to
    solitonic solutions. The set of asymptotic regularity requirements
    (\ref{as1}), (\ref{mu}) for self-gravitating solutions is thus much
    stronger than (\ref{Efin}) and contains the latter as a corollary.

    Specifically, if we use the harmonic $u$ coordinate, it is easily shown
    that both a regular axis and a regular spatial asymptotic require $u\to
    \pm \infty$. Choosing, without loss of generality, $u_{\rm ax}= -\infty$
    and $u_\infty = +\infty$, one finds that at a regular axis, in addition
    to (\ref{ax1}) and (\ref{ax2}),
\beq
     r = \e^\beta \sim \e^{cu},\quad
     c = \e^{\gamma+\xi}\Big|_{u\to -\infty}  = \const > 0.  \label{ax3}
\eeq
    At a regular spatial asymptotic ($u\to +\infty$), in addition to
    (\ref{as1}) and (\ref{mu}), we have
\beq
    \e^{\alpha} \approx \e^\beta = r \sim \e^{(1-\mu)u},     \label{as3}
\eeq
    and the SET components $T\mN$ must decay at large $u$ quicker than
    $r^{-2} \sim \e^{-2(1-\mu)u}$.

\section {Radial electromagnetic fields}

    The Einstein-Maxwell equations for a radial electromagnetic field lead
    to the metric \cite{Bron3}
\bearr
     ds^2 = \frac{K dt^2}{s^2 (h, u)} - \frac{s^2 (h, u)}{K}
            \bigl[ \e^{2(a+b)u} du^2
\nnn \inch\cm
    + \e^{2a u} dz^2 \bigr] + \e^{2b u} d \varphi^2,     \label{dsR1}
\ear
    where $K = (G q^2)^{-1}$, $q^2 = q_e^2 + q_m^2$,
    $a,\ b =\const$, the function $s(h,u)$ is defined as
\bear
    s(h,u)= \vars{
             h^{-1} \sinh (h u), \ &  \ \ h > 0,    \\
                              u, \ &  \ \ h = 0,    \\
               h^{-1}\sin (h u), \ &  \ \ h < 0,    }
\ear
    and $h^2 \sign h = ab$. The electromagnetic field is represented by
\beq
     F^{01} = q_e \e^{-2\alpha},\cm   F_{23} = q_m,      \label{emR1}
\eeq
    the constants $q_e$ and $q_m$ being the linear electric and magnetic
    charges, respectively.

    This space-time possesses a singular axis (charged thread) and spatial
    infinity if $h\geq 0$ and $h+B > 0$. The large $r$ asymptotic then
    cannot be regular, in particular, $g_{tt} \to 0$, and
    $|\beta'|\e^{\beta-\alpha} \sim \e^{-au}$ does not tend to a finite
    constant. In other cases we have two singular axes.

    Our more general \cyl\ system with nonlinear radial electric and
    magnetic fields closely resembles a system with the same Lagrangian
    (\ref{L}) in the \sph\ case \cite{Bron1}. It has been shown in
    Ref.\,\cite{Bron1} that the field system (\ref{L}), with any function
    $\Phi (F)$  having a Maxwell asymptotic at small $F$, does not admit a
    \ssph\ solution with a regular centre and a nonzero electric charge. We
    shall see that a similar theorem can be proved using the same arguments
    for radial fields and the metric (\ref{ds}). Moreover, {\sl both here
    and in the \sph\ case}, instead of requiring a Maxwell asymptotic,
    {\sl it is sufficient to assume that the function $\Phi(F)$ is simply
    smooth at $F=0$}.

    Indeed, \eqs (\ref{eqF}) and (\ref{eqF*}) give
\bear
    F^{01} \Phi_F = q_e \e^{-2\alpha}, \cm
    F_{23} = q_m,                                            \label{F0123}
\ear
    where, as before, the constants $q_e$ and $q_m$ are the linear electric
    and magnetic charges. It follows that
\bear
     E^2 \Phi_F ^2 \eql q_e ^2 \e^{-2\xi -2\beta},            \label{Erad}
\\
     B^2 \eql     q_m ^2 \e^{-2\xi -2\beta},                  \label{Brad}
\ear
     and the nonzero SET components are
\bearr
     T_0 ^0 = T_1 ^1 = \fpi (4E^2 \Phi_F + \Phi),        \label{01R}
\\ \lal
     T_2 ^2 = T_3 ^2 = \fpi (-4B^2 \Phi_F + \Phi).       \label{23R}
\ear

\Theorem {Theorem 1}  {The system (\ref{L}), such that $\Phi (F)$ is
     $C^1$-smooth at $F=0$, does not admit a static, \cyl\ solution for
     radial electromagnetic fields with a regular axis and a nonzero
    electric charge.}

\noi
     {\bf Proof}. From  the condition $T\mN < \infty$ it follows
\beq
      (E^2 +B^2)|\Phi_F| < \infty.                           \label{th1}
\eeq

     Suppose first that $q_m =0$, $q_e \neq 0$. Therefore, from (\ref{Erad})
     and (\ref{th1}) it follows that, at a regular axis, $E^2 \Phi_F$ is
     finite whereas $E^2 \Phi_F^2 \to \infty$. These conditions combined
     lead to $F\to 0$ and $\Phi_F \to \infty$, that is, a singular
     behaviour of $\Phi$ at small  $F$. Thus for purely electric fields the
     theorem is valid.

     Suppose now $q_m \neq0$ and $q_e \neq 0$. Then (\ref{th1}) should hold
     for $E^2$ and $B^2$ taken separately. As stated previously, this
     condition applied to $E^2$, combined with (\ref{Erad}), leads to
     $\Phi_F \to \infty$. But $B^2$ also tends to infinity, so even stronger
     $B^2 \Phi_F \to \infty$, violating (\ref{th1}). The theorem is proved.

\medskip
     Consider the remaining case of a purely magnetic radial
     field. Then, according to (\ref{01R}),
\beq
     T_0^0 = T_1^1 = \Phi/(16\pi).                           \label{01R'}
\eeq
     Meanwhile, by \eq (\ref{Brad}), $F = 2B^2 \to \infty$ at a possible
     regular axis where $\xi\to\const$ and $\beta\to-\infty$, therefore the
     necessary condition (\ref{Tfin}) for regularity on the axis holds as
     long as $\Phi \to \const$ as $F\to +\infty$. Moreover, it then follows
     that $\Phi_F = o(1/F)$, hence $B^2 \Phi_F \to 0$, which means that the
     SET (\ref{01R}), (\ref{23R}) has the structure of a cosmological
     term: $T\mN \sim \Phi\delta\mN$ near a regular axis. We conclude that
     {\sl a possible solution with a radial magnetic field, possessing a
     regular axis, is approximately de Sitter or anti-de Sitter near such an
     axis}.

     As for a regular spatial asymptotic, we have seen that it is absent in
     the Einstein-Maxwell solution (except for the trivial case of flat
     space and zero $F\mn$). The same will evidently be true if $\Phi(F)$
     has a Maxwell behaviour, $\Phi(F) \sim F$, at small $F$: the
     NED-Einstein solution will then behave at large $r$ as an
     Einstein-Maxwell solution.

     As follows from (\ref{01R}) and (\ref{23R}), a regular spatial
     asymptotic requires $F\Phi_F = o(1/r^2)$ at large $r$. For a purely
     magnetic solution, in case $\xi\to \const$, we have $F\sim 1/r^2$,
     hence $\Phi_F = o(1)$. This is a necessary condition of regularity.
     Another evident necessary condition is $\Phi = o(1/r^2)$.

     Let us return to the Einstein equations. Due to $T_2^2 = T_3^3$, they
     lead to
\beq
     \beta'' - \xi'' = 0 \then
     \beta (u) = \xi (u) + c_1 u,                           \label{beta-R}
\eeq
     where $c_1 = \const$ and another integration constant is set equal to
     zero by choosing the scale along the $z$ axis. To have a regular axis
     at $u\to -\infty$, we should put $c_1 > 0$.

     A sum of (\ref{00}) and (\ref{11}) leads to
\beq
     2\beta'' = -G (4E^2 \Phi_F +\Phi) \e^{2\alpha}.    \label{beta-fi}
\eeq
     We can also obtain a relation between $\beta (u)$ and $\gamma (u)$:
     the difference of (\ref{00}) and (\ref{11}) gives
\beq
     \beta'' - \beta'^2 + \beta' c_1 + \gamma' c_1 - 2 \beta'\gamma'=0.
                                                        \label{bg-R}
\eeq

     It is hard to obtain an exact solution to the NED-Einstein equations
     with a given function $\Phi(F)$. However, for a purely magnetic
     field, the general solution can be described by specifying $\beta (u)$
     if we consider $\Phi(F)$ as one of the unknown functions. Indeed, one
     then finds $\xi(u)$ and $\gamma(u)$ from (\ref{beta-R}) and
     (\ref{bg-R}), then $\alpha (u)$ from (\ref{harm}), $\Phi (u)$
     from (\ref{beta-fi}) and $F(u)$ from (\ref{Brad}). A comparison of the
     latter two functions leads to $\Phi(F)$.

     We conclude that solitonic solutions, regular both on the axis and at
     infinity, are not excluded with radial magnetic fields. Necessary
     conditions for obtaining such a solution are: (i) $q_e=0$, $q_m\ne 0$;
     (ii) $\Phi(F) \to \const$ as $F\to +\infty$ and (iii)
     $\Phi = o(F)$ as $F\to 0$ (a non-Maxwell behaviour at small $F$).

     One more necessary condition follows from (\ref{beta-R}) and
     (\ref{beta-fi}). Namely, due to (\ref{beta-R}), $\beta' = c_1$
     both on the axis and on the asymptotic, and integration of
     (\ref{beta-fi}) gives:
\beq
      \int_{ -\infty}^{+\infty} \Phi \e^{2\alpha}du =0,  \label{int-R}
\eeq
     whence it follows that $\Phi (F)$ should have an alternating sign in
     the range $F = 2B^2 > 0$ corresponding to $u\in\R$.

     It should be stressed that all these conditions are only necessary.
     Even if all of them hold, one cannot guarantee that a given function
     $\beta(u)$ will lead to a valid solitonic solution. Thus, a nontrivial
     requirement is that the resulting $\Phi(F)$ should be a function. In
     particular, if the function $F = 2B^2(u)$ obtained from (\ref{Brad}) is
     monotonic, then $\Phi(F)$ will be a function only if $\Phi(u)$ obtained
     from (\ref{beta-fi}) (with $E=0$) is also monotonic. This in turn
     requires (according to the Einstein equations) that the difference
     $\beta'' - \gamma''$ should have an invariable sign at all $u$.

\section {Azimuthal electromagnetic fields}

     The NED-Einstein equations for azimuthal electromagnetic fields
     can be studied in quite a similar manner as for radial fields.
     We therefore only mention the main points, avoiding the details.

     The Einstein-Maxwell equations for an azimuthal electromagnetic field
     lead to the metric \cite{Bron3}
\bearr
     ds^2 = \frac{\cosh^2 (hu)}{Kh^2} \bigl[\e^{2bu}d t^2 - \e^{2(a+b)u}du^2
\nnn \inch
     - \e^{2au} d \varphi^2 \bigr] - \frac{K h^2}{\cosh^2 (hu)}   dz^2,
                                                        \label{dsA1}
\ear
     where $K = [G(i_e^2 + i_m^2)]^{-1} $, $h^2 = ab$, $a, b = \const$,
     $a>0$, $b>0$, and the electromagnetic field given by
\beq
      F_{03} = i_m = \const; \quad\                     \label{FA1}
      F^{12} = i_e \e^{-2\alpha}, \ \ i_e = \const,
\eeq
     where $i_e$ and $i_m$ are the effective currents of electric and
     magnetic charges along the $z$ axis, respectively.

     This solution, like (\ref{dsR1}), (\ref{emR1}), does not provide a
     regular axis or a regular large $r$ asymptotic for any values of the
     integration constants.

     In NED under consideration, \eqs (\ref{eqF}) and (\ref{eqF*}) give
\beq
     F_{03} = i_m, \cm
     F^{12}\Phi_F = i_e \e^{-2\alpha},                    \label{F_A}
\eeq
     which leads to
\bear
     E^2 \eql i_m^2 \e^{-2\gamma-2\beta},                     \label{E_A}
\\
     B^2 \Phi_F^2 \eql i_e^2 \e^{-2\gamma -2\beta}.           \label{B_A}
\ear
     The SET components have the form
\bear
      T_0^0 = T_3^3 \eql \fpi(4 E^2 \Phi_F + \Phi),           \label{03A}
\\
      T_1^1 = T_2^2 \eql \fpi (-4B^2 \Phi_F + \Phi).          \label{12A}
\ear

     Using the same arguments as in the case of radial fields, we can obtain
     an analogue of Theorem 1 for azimuthal fields:

\Theorem {Theorem 2} {The system (\ref{L}), such that $\Phi (F)$ is
     $C^1$-smooth at $F=0$, does not admit a static, \cyl\ solution for
     azimuthal electromagnetic fields with a regular axis and a nonzero
     effective electric current $i_e$.}

     Thus a regular axis is not excluded only in the case of an azimuthal
     electric field, which can result from an effective magnetic current.

     All this has been obtained without using the Einstein equations. The
     latter give, in full similarity with \sect 3:
\beq
     \beta (u) = \gamma (u) + c_2 u, \ \ c_2 = \const,     \label{beta-A}
\eeq
     with $c_2 > 0$ if we wish to have a regular axis;
\beq
     2\beta '' = G(4B^2 \Phi_F -\Phi) \e^{2\alpha},      \label{beta''-A}
\eeq
     and
\beq
     \beta'' - \beta^{'2} + \beta' c_2 +\xi' c_2 - 2\beta' \xi' =0.
                                                         \label{bx-A}
\eeq

     It is again hard to obtain an exact solution with a given function
     $\Phi(F)$. However, in case $i_e=0$, the general solution may be
     parametrized by $\beta (u)$. Indeed, given $\beta(u)$, we can find
     $\gamma(u)$ and $\xi(u)$ from (\ref{beta-A}) and (\ref{bx-A}),
     then $\alpha (u)$ from (\ref{harm}), $\Phi (u)$ from (\ref{beta''-A})
     and $F(u)$ from (\ref{E_A}). Comparing the latter two functions, we can
     find $\Phi(F)$.

     A regular asymptotic again needs a non-Maxwell behaviour of $\Phi(F)$
     at small $F$: $\Phi = o(F)$.

     We conclude that solitonic solutions, regular both on the axis and at
     infinity, are not excluded with azimuthal electric fields. Necessary
     conditions for obtaining such a solution are: (i) $i_e=0$, $i_m\ne 0$;
     (ii) $\Phi(F) \to \const$ as $F\to -\infty$, (iii) $\Phi = o(F)$ as
     $F\to 0$, and (iv) an alternating sign of $\Phi(F)$ in the range $F =
     -2E^2 < 0$ corresponding to $u\in\R$. Again, as described above for
     R-fields, these conditions are not sufficient, and, in particular, a
     nontrivial problem is to provide that the resulting $\Phi(F)$ will be a
     function.

\section {Longitudinal electromagnetic fields}

     The Einstein-Maxwell equations for a longitudinal electromagnetic field
     lead to the metric \cite{Bron3}
\bear
     ds^2 = \frac{\cosh^2 (h u)}{Kh^2}\bigl[\e^{2bu}d t^2 - \e^{2(a+b)u}du^2
\nn
     - \e^{2au} dz^2 \bigr] - \frac{Kh^2}{\cosh^2 (hu)} d\varphi^2,
                                                          \label{dsL1}
\ear
     where $K = [G(i_e^2 + i_m^2)]^{-1} $, $h^2 = ab$, $a,b = \const$,
     $a>0$, $b>0$, and the electromagnetic field is given by
\beq
      F_{02} = i_m = \const; \quad\                       \label{FL1}
      F^{13} = i_e \e^{-2\alpha}, \ \ i_e = \const,
\eeq
     where $i_e$ and $i_m$ are the effective solinoidal electric and
     magnetic charges, respectively.

     The metric (\ref{dsL1}) does not admit a spatial infinity since
     $g_{\varphi\varphi}$ is bounded above. Instead, there are two axes
     at $u\to\pm \infty$. The case $a=b=h$, $Kh^3=1$ corresponds to Melvin's
     nonsingular universe \cite{Melvin}, where $u\to -\infty$ is a regular
     axis while the other axis, $u=+\infty$, is infinitely remote.

     In our NED-Einstein system, \eqs (\ref{eqF}) and (\ref{eqF*}) give
\beq
     F_{02}=i_m, \qquad
     F^{13} \Phi _F = i_e \e^{-2\alpha},              \label{F_L}
\eeq
     and
\bear
     E^2 \eql i_m^2 \e^{-2\gamma -2\xi},                     \label{E_L}
\\
     B^2 \Phi_F^2 \eql i_e^2 \e^{-2\gamma -2\xi}.            \label{B_L}
\ear
     The SET components are
\bear
     T_0^0 = T_2^2 \eql \fpi(4E^2 \Phi_F + \Phi),            \label{02L}
\\
     T_1^1 = T_3^3 \eql \fpi(-4B^2 \Phi_F + \Phi).           \label{13L}
\ear

     The above Einstein-Maxwell solution shows that a regular axis in an
     L-field even exists with a linear electromagnetic field, therefore in
     NED it can also be easily obtained: if only the derivative $\Phi_F$
     is finite at the corresponding value of $F$, a regular axis is
     obtained just as with the Maxwell field. Furthermore, assuming that
     there is a regular axis, the Einstein equations can be solved by
     quadratures for an arbitrary choice of $\Phi(F)$.

     Indeed, the difference of \eqs (\ref{00}) and (\ref{22}) leads to
\beq
     \xi (u) = \gamma (u) + c_3 u, \cm c_3 = \const,     \label{L-xi}
\eeq
     and to have a regular axis at $u\to-\infty$ we should put $c_3 =0$.
     Then, the difference of \eqs (\ref{11}) and (\ref{22}) gives
\beq
     \xi'' - \xi'^2  - 2\beta' \xi' =0,                  \label{L-b}
\eeq
     whence
\beq
     C\e^{2\beta}= \e^{-\xi} \xi' , \cm C=\const.        \label{L-b1}
\eeq
     Then, provided $C \neq 0$, \eq (\ref{11}) with (\ref{13L}) leads to
\beq
     C\xi'' = - \kappa T_1^1 \e^{3\xi} \xi' .            \label{L-xi1}
\eeq

     Now, due to $\xi=\gamma$, the SET components can be obtained
     as functions of $\xi$: since
\[
     F = 2(B^2-E^2) = 2\e^{-4\xi} [i_e^2 \Phi_F^{-2} - i_m^2],
\]
     $\xi$ is expressed in terms of $F$ when $\Phi(F)$ is specified. Then
     $F$ and $\Phi(F)$, and hence $T_1^1$, are determined as functions of
     $\xi$.  As a result, \eq (\ref{L-xi1}) can be solved by quadratures:
\beq
     u =-\int\frac{C d\xi}{\kappa}
        \biggl(\int T_1^1 \e^{3\xi}d \xi\biggr)^{-1}.    \label{int-L}
\eeq

     Regularity on the axis $u=-\infty$ can be achieved by adjusting
     the integration constants as long as $\Phi(F)$ is a smooth function
     since, with finite $\xi=\gamma$ and $\Phi_F$, the strengths $E$ and $B$
     are finite according to (\ref{E_L}) and (\ref{B_L}).

     It turns out, however, that the above solution with $\xi = \gamma$
     cannot possess a regular spatial asymptotic. Indeed, as follows from
     (\ref{L-b1}), $(\e^{-\xi})' \to \infty$ as $\beta \to \infty$,
     whereas the requirement (\ref{as1}) leads to $(\e^{-\xi})'\to 0$.
     In particular, solutions (\ref{L-xi}), (\ref{L-b1}) with $C\ne 0$,
     having a regular axis, cannot have a regular asymptotic.

     It remains to consider the special case $C=0$ in (\ref{L-b1}), when we
     can put $\xi \equiv \gamma \equiv 0$ without loss of generality. Then
     \eqs (\ref{E_L}) and (\ref{B_L}), leading to
\beq
     E^2 = i_m^2,\cm   B^2 \Phi_F^2 = i_e^2,              \label{Ls1}
\eeq
     make it possible to determine the function $\Phi(F)$:
\beq
    \Phi^2 (F) = 8 i_e^2 (F + 2i_m^2),                    \label{Ls2}
\eeq
     with an evidently non-Maxwell behaviour at small $F$. Now, the
     Einstein equations (\ref{11}) and (\ref{33}) lead to $T^1_1 =
     T^3_3 \equiv 0$, and one can easily verify that this condition is
     satisfied by $\Phi (F)$ taken according to \eq (\ref{Ls2}). The only
     remaining field equation to be solved is
\beq
     \e^{-2\beta} \beta'' = -\frac{4Gi_e^2}{\Phi}(F + 4i_m^2), \label{Ls3}
\eeq
     which follows from (\ref{00}) or (\ref{22}). \eq (\ref{Ls3}) connects
     two unknown functions $F(u)$ and $\beta(u)$, so that one function may
     be chosen arbitrarily.

     This choice makes it possible to obtain solitonic solutions, but only
     with a purely magnetic field created by the effective current $i_e$
     (one can check that a solution with $i_m \neq 0$ cannot have a regular
     asymptotic). We now have from (\ref{Ls3})
\beq
     G\Phi = \pm G\sqrt{8 i_e^2 F} = -2\beta''\e^{-2\beta}.     \label{Ls4}
\eeq
     Fixing the sign on the l.h.s., we obtain that $\beta''$ has the same
     sign at all $u$, hence, in a nontrivial solution, $\beta'(+\infty) \ne
     \beta'(-\infty)$. With $\gamma\equiv \xi\equiv 0$, a regular axis
     corresponds to $\beta'(-\infty)=1$, and, at a regular asymptotic,
     $\beta'(+\infty) = 1-\mu$. Therefore, (i) a nontrivial solution
     inevitably leads to $\mu \ne 0$ and (ii) the angular defect $\mu >0$ is
     obtained by choosing $+$ in \eq (\ref{Ls4}).

     Let us give a particular example of a solitonic solution belonging to
     this exceptional family. We choose
\beq
     \e^{2\beta} = r_0^2 \e^{(2-\mu)u} (\cosh ku)^{-\mu/k},    \label{Ls5}
\eeq
     with the constants $r_0$ (an arbitrary length), $\mu < 1$ (the angular
     defect) and $k \geq 1$ (which is required by the regularity condition
     (\ref{ax2})). Then \eq (\ref{Ls4}) gives
\beq                                                            \label{Ls6}
     G\sqrt{8i_e^2 F} = \frac{k\mu}{2r_0^2} \e^{(\mu-2)u}
                        (\cosh ku)^{-2-\mu/k}.
\eeq
     As is easily verified, $F$ turns to zero as $u\to \pm \infty$, and the
     sufficient conditions of regularity are satisfied.

\section {Concluding remarks}

     We have considered static cylindrically symmetric NED-Einstein
     equations with an arbitrary gauge-invariant NED Lagrangian of the form
     $\Phi(F)$.

     For R-fields we have proved a theorem that a regular axis is impossible
     if the electric charge is nonzero and the function $\Phi (F)$ is
     regular at $F=0$. A similar theorem for A-fields tells us that a
     regular axis is impossible if there is a nonzero effective longitudinal
     electric current and $\Phi(F)$ is regular at $F=0$. For both R- and
     A-fields, regular stringlike solutions are not excluded; they should
     possess a nonzero effective magnetic charge or a nonzero effective
     longitudinal magnetic current, respectively. Necessary conditions for
     the existence of such solutions are, among others, (i) a non-Maxwell
     behaviour of $\Phi(F)$ at small $F$ and (ii) an alternating sign of
     $\Phi(F)$.

     For L-fields, there exist configurations having a regular axis, and a
     general exact solution for such configurations was found by quadratures:
     specifying $\Phi (F)$, we can obtain all metric functions. In other
     cases, we have found parametrizations of the general solution in terms
     of one of the metric functions: knowing it, we can find other metric
     functions and $\Phi (F)$. A certain problem is then to obtain $\Phi(F)$
     as a (single-valued) function since $\Phi$ and $F$ are found separately
     as functions of the radial coordinate. It is for this reason that we
     do not give any explicit examples of solitonic solutions for R- and
     A-fields.

     We have found that for L-fields, in addition to the above general
     solution, there is an exceptional solution ($C=0$ in \eq (\ref{L-b1}))
     for a special choice (\ref{Ls2}) of $\Phi(F)$. Its properties are
     drastically different from those of the general solution, and, in
     particular, it gives rise to purely magnetic solitonic configurations
     with a nonzero angular defect $\mu$. A specific example of such a
     configuration is given by \eqs (\ref{Ls5}), (\ref{Ls6}).

     In the present study, just as in the case of spherical symmetry
     \cite{Bron1}, we find that the properties of electric and magnetic
     fields are quite different. In particular, only purely magnetic
     R-fields and purely electric A-fields can form regular self-gravitating
     configurations. This is a clear manifestation of the absence, in
     nonlinear theory, of the electric-magnetic duality which is so
     important in the Maxwell theory. It can be of interest to study another
     symmetry, the so-called $FP$ duality \cite{Bron1} between different
     theories of NED, for various \cyl\ configurations; this, however, goes
     beyond the scope of this paper.

\end{document}